\documentclass[12pt,preprint]{aastex}
\renewcommand{\cite}{\citealp}

 \usepackage{emulateapj5}
   
  \usepackage{apjfonts}

\newcommand{\realfigure}[3]{ 
             \hbox{~} \centerline{\includegraphics[width=8.7cm]{#1}}
             \figcaption{#2 \label{#3}} \vspace{0.05in}\centerline{}}

\newcommand{\letter}{{\it Letter}}
\newcommand{\feh}{{\rm [Fe/H]}}

\newcommand{\mainpaper}{{E. V. Held et al. 2003, in prep.}}
\newcommand{\abbrev}[1]{{#1}}

\newcommand{\mscred}{{\sc mscred}}
\newcommand{\wfpred}{{\sc wfpred}}

\newcommand{\caii}{Ca\,{\sc ii}\ }   

\begin{document}

\title{Clues on the evolution of the Carina dwarf spheroidal 
galaxy from the color distribution of its red giant stars
\altaffilmark{1}}

\author{
Luca Rizzi,\altaffilmark{2}$^{,}$\altaffilmark{3}
Enrico V. Held,\altaffilmark{2}
Gianpaolo Bertelli,\altaffilmark{4}$^{,}$\altaffilmark{2}
Ivo Saviane\altaffilmark{5}
}

\altaffiltext{1}{Based on data collected at E.S.O. La Silla, Chile, 
Proposals~No.~63.N-0017 and 64.N--0512}

\altaffiltext{2}{INAF, Osservatorio Astronomico di
Padova, vicolo dell'Osservatorio 5, I-35122 Padova, Italy;
(held,rizzi)@pd.astro.it}

\altaffiltext{3}{Dipartimento di Astronomia, Universit\`a di 
Padova, vicolo dell'Osservatorio 2, I-35122 Padova, Italy}

\altaffiltext{4}{Istituto di Astrofisica Spaziale e Fisica Cosmica,
CNR, Area di Ricerca Tor Vergata, Via del Fosso del Cavaliere 100,
00133 Roma, Italy}

\altaffiltext{5}{European Southern Observatory, Casilla 19001,
Santiago 19, Chile}

\begin{abstract}
The thin red giant branch (RGB) of the Carina dwarf spheroidal galaxy
appears at first sight quite puzzling and seemingly in contrast with
the presence of several distinct bursts of star formation. In this
\letter, we provide a measurement of the color spread of red
giant stars in Carina based on new $BVI$ wide-field observations, and
model the width of the RGB by means of synthetic color-magnitude
diagrams.
The measured color spread, $\sigma_{V-I}=0.021 \pm 0.005$, is quite
naturally accounted for by the star-formation history of the galaxy.
The thin RGB appears to be essentially related to the limited age range 
of its dominant stellar populations, with no need for a 
metallicity dispersion at
a given age. This result is relatively robust with respect to changes 
in the assumed age-metallicity relation, as long as the mean
metallicity over the galaxy lifetime matches the observed value
([Fe/H]$=-1.91 \pm 0.12$ after correction for the age effects).
This analysis of photometric data also sets some constraints on the
chemical evolution of Carina by indicating that the chemical abundance
of the interstellar medium in Carina remained low
throughout each episode of star formation even though these
episodes occurred over many Gyr.
\end{abstract}


\section{Introduction}

Since its discovery in 1977 (Cannon, Hawarder, \& Tritton
\cite{can+77}), the Carina dwarf spheroidal (\abbrev{dSph}) galaxy has
played an important role in our understanding of the nature of dwarf
galaxies. Carina was in fact the first dSph galaxy in which a
prominent intermediate age population was recognized (Mould \&
Aaronson \cite{mou+aar83}).
The presence of a non-negligible component of old stars was
subsequently revealed by the detection of a large number of RR~Lyrae
variable stars  (Saha, Monet, \& Seitzer \cite{sah+86}; 
Dall'Ora et al. \cite{dall+03}).
Mighell (\cite{migh90}) estimated the old population to constitute
$(17 \pm 4) \%$ of the galaxy population, a number confirmed by
Smecker-Hane et al. (\cite{smec+94}) and Mighell (\cite{migh97}).
A detailed description of the star-formation history (\abbrev{SFH}) of
Carina was given by Smecker-Hane et al. (\cite{smec+94},
\cite{smec+96}).  A deep $BR$ color-magnitude diagram (\abbrev{CMD})
revealed the presence of three distinct turn-offs, pointing at three
separate episodes of star formation approximately $2$, $3-6$ and
$11-13$ Gyr ago {(see also Monelli et al. \cite{mone+03})}. 
A quantitative assessment of this scenario was then
provided by CMD simulation techniques applied to CTIO 4m and HST data
(Hurley-Keller, Mateo, \& Nemec \cite{hurl+98}; Hernandez, Gilmore, \&
Valls-Gabaud \cite{her+00}; Dolphin \cite{dol02}).

Quite surprisingly given this complex star-formation history, 
Carina shows a
{\it very thin red giant branch} with all the different turn-offs
connecting at the RGB base (Smecker-Hane et al. \cite{smec+94},
\cite{smec+96}).  This seems to imply that either no significant metal
enrichment took place in Carina, or that age and chemical enrichment
of the stellar populations played in opposite directions to yield a
narrow RGB.  While these alternative views have often been discussed,
they have never been quantitatively addressed.
%
%
The chemical evolution of dwarf spheroidal galaxies, in particular the
role of galaxy-wide gas outflows and/or infall, remains a
controversial issue and many basic questions are still open.  Recent
efforts to model the chemical evolution of dwarf spheroidal galaxies
have been presented by Ikuta \& Arimoto (\cite{ikut+arim02}), Carigi,
Hernandez, \& Gilmore (\cite{cari+02}), and Tolstoy et
al. (\cite{tols+03}). All groups use known SFHs to infer the 
chemical-enrichment 
histories of dSph galaxies, but the scenarios differ in the
role played by galactic winds and gas infall.

This \letter\ shows that realistic modeling of the color distribution
of red giant stars by population synthesis methods can help
constraining the chemical evolution of dwarf spheroidals galaxies.
The intrinsic color width of the upper RGB in Carina is quantified here 
by exploiting the high statistics of a new $BVI$ database
from wide-field observations.  The star-formation history of Carina is
then derived for a set of different metal-enrichment histories (the
age-metallicity relation of this galaxy is still uncertain) by
comparing observed and synthetic color-magnitude diagrams.  The narrow
RGB of Carina is shown to be consistent with the star-formation
history derived from the observed CMDs, under reasonable assumptions
on the chemical-enrichment history.

\section{The data}

Observations of Carina were carried out using the
{\it Wide Field Imager} at the ESO/MPG 2.2m telescope. The
camera consists of a mosaic of eight 2k $\times$ 4k EEV CCDs, in a
square array of 2 $\times$ 4 detectors. In this configuration, the
total number of pixels is 8k $\times$ 8k, with a
scale of 0.238 arcsec pixel$^{-1}$ and a field of view 
$32\arcmin \times 32\arcmin$. Three different fields were observed: a
central field and two outer fields with
$\sim 20\arcmin$ offsets. Sky conditions were photometric, and
standard star fields taken from the list of Landolt (\cite{lan92})
were observed for calibration purpose.
A full description of the data and reduction methods will be presented
in a forthcoming paper (\mainpaper) while just a
brief outline is given here.  Pre-reduction of CCD
images was performed within the IRAF environment.  Each image was
bias-subtracted and flat-fielded using twilight sky flats. After these
steps, all images were astrometrically calibrated using the IRAF
\mscred\ package (Valdes \cite{vald98}) and the script package
\wfpred\ developed at the Padova Observatory. 
%

Photometry was obtained using {\sc daophot~ii/allstar} (Stetson
\cite{ste87}) on the co-added images, 
and the results were calibrated using 
calibration relations computed from standard stars observed in
all CCDs. Similarly, aperture corrections were computed using a
growth curve analysis performed on bright isolated stars in each CCD.
%
%
Artificial star tests were done 
to evaluate the uncertainty of photometry and the degree of
completeness of the data.
%
%
The photometric errors in the $V$ band range from $\sim 0.01$ mag near
the tip of the RGB, to $\sim 0.3$ mag near the detection limit.


\section{Modeling the red giant branch of Carina}

\subsection{RGB width measurement}
\label{sec:larghezza}

These wide field data allowed us to obtain a sound 
quantitative estimate of the RGB width in Carina. In order to measure
the color spread of the RGB, a fiducial ridge line was fitted to the
mean colors of red giants in $0.5$ mag bins.  The distribution of
color residuals about the fiducial line was then measured down to 1.5
mag below the tip.  After quadratic subtraction of the contribution of
photometric errors, the estimated intrinsic color spread of the upper
RGB is:
\begin{center}
$\sigma_{(V-I)} = 0.021 \pm 0.005 $
\end{center}
where the statistical error was estimated by bootstrap resampling of the
RGB sample in 500 trials.

Using a direct comparison of the red giant branch of Carina with the
ridge lines of Galactic globular clusters (Da Costa \& Armandroff
\cite{dac+arm90}), we estimated a mean metallicity ${\rm [Fe/H]} =
-2.08 \pm 0.12$ on the Zinn \& West (\cite{zinn+west84}) scale
(\mainpaper).
This value agrees well with previous photometric determinations (Mould
\& Aaronson \cite{mou+aar83}; Mighell \cite{migh90}; Smecker-Hane et
al. \cite{smec+94}; Mateo, Hurley-Keller, \& Nemec \cite{mat+hur98}),
{as well as the low dispersion spectroscopic 
measurements of the \caii\ lines of 
Da Costa (\cite{dac94})
and Smecker-Hane et al. (\cite{smec+99}).  
}
A somewhat higher metallicity, ${\rm [Fe/H]} = -1.6$ with a spread of
0.5 dex, has recently been derived by Tolstoy et al. (\cite{tols+03})
from high-resolution spectroscopy of 5 red giant stars.

The difference between this new spectroscopic estimate and
previous photometric [Fe/H] value can be only partly accounted for
by age effects on the RGB color.  The stars in Carina are, in fact,
$\sim 7$ Gyr younger than the stellar populations of Galactic globular
clusters (see below), and we know that younger isochrones are shifted
blueward (see, e.g., Da Costa \cite{daco98}).
Using Girardi's et al. (\cite{gira+00}) models in the range $Z=0.0001$
to 0.0004, we find the effects of a younger age on the $(V-I)$ color
to be $\sim 0.035$ mag at the RGB level used for metallicity
estimation. This bluer color mimics a lower metallicity, by $\sim
0.17$ dex (see also Smecker-Hane et al. \cite{smec+94}). 
By taking this effect into account, the age-corrected
metallicity of Carina is then [Fe/H]$=-1.91 \pm 0.12$.

\realfigure{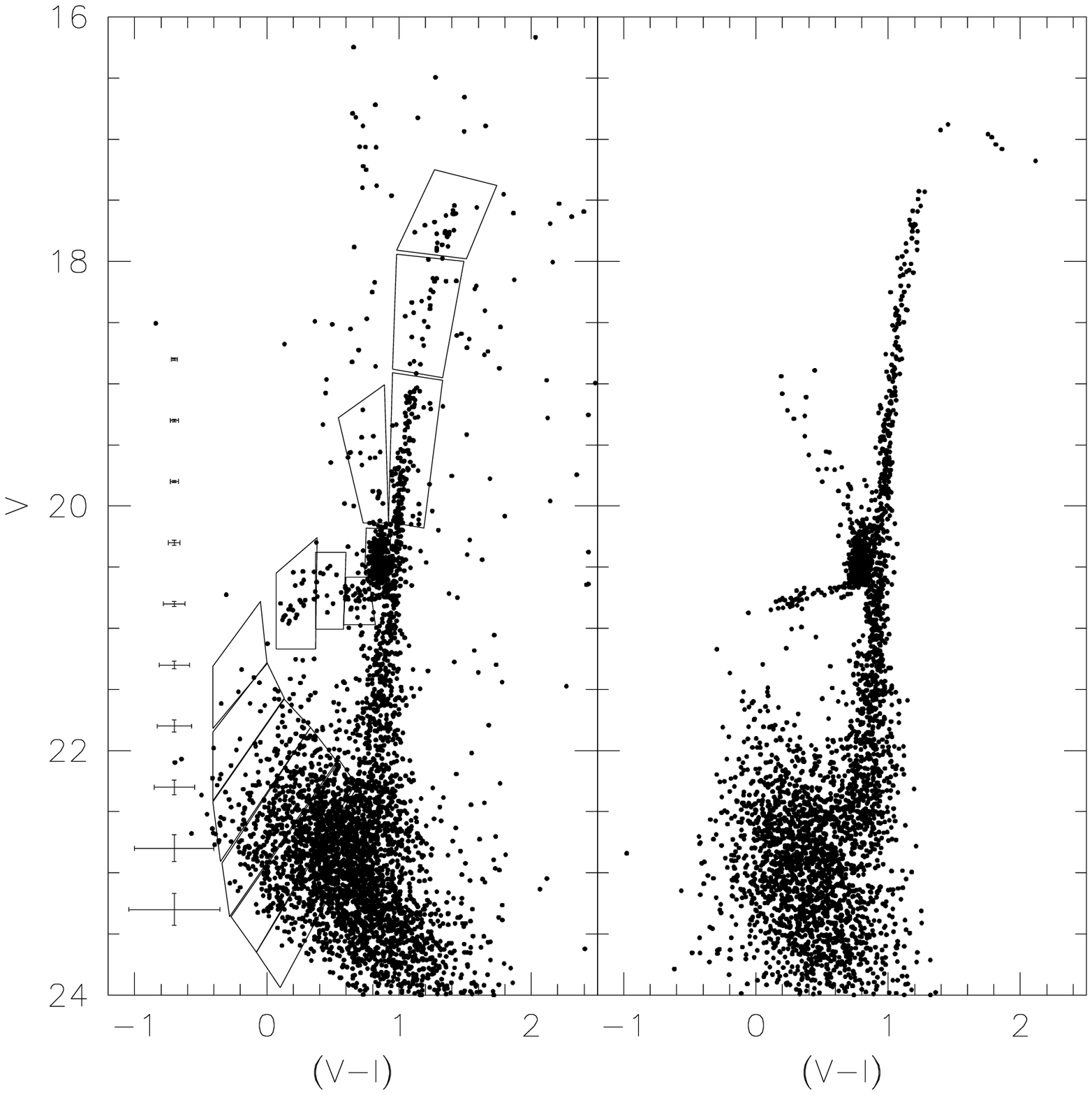}{
Comparison of the observed and synthetic color-magnitude diagrams for
Carina.  The {\it left} panel shows the statistically decontaminated
CMD of the central region of Carina, along with the boxes used by the
$\chi^2$ minimization technique. The {\it right} panel shows the
simulated color-magnitude diagram, assuming the PT law described 
in the text. 
}{f_dati}

The color distribution of RGB stars formally corresponds (for an old
population) to a metallicity dispersion $\sigma_{\rm[Fe/H]}=0.10 \pm
0.04$, which is significantly smaller than the quoted dispersion of
spectroscopic measurements. 

\subsection{Population synthesis}


Figure~\ref{f_dati} ({left panel}) shows the CMD of the the central
part of Carina, for which simulations were performed. 
An elliptical region with a 12 arcmin semi-major axis, $65^{\circ}$
position angle, and $0.33$ ellipticity (Irwin \& Hatzidimitriou
\cite{irw+hat95}) was selected to this purpose.  The intrinsically
poor statistics of the red giant stars in Carina prevents reliable
measurements of the RGB width in the outer regions.
The observed CMD was simulated using a parametric technique similar to
that employed by Aparicio, Gallart, \& Bertelli (\cite{apa+97a}), and
a description of our methods is presented by Rizzi et
al. (\cite{riz+01}).  In brief, our technique assumes a number of
known input parameters, such as the initial mass function (IMF), the
distance to the galaxy, and the reddening.  A Salpeter (\cite{sal55})
IMF is adopted, since recent work suggests a general validity of this
IMF for masses larger than $0.5\, M_{\odot}$ (Scalo \cite{sca98};
Kroupa \cite{kro00}). A distance modulus $(m-M)_0=20.00 \pm 0.06$ and
a reddening $E(B-V)=0.045 \pm 0.01$ were adopted (\mainpaper).
Using these input parameters, a set of synthetic stellar populations
were generated using models from the library of Girardi et
al. (\cite{gira+00}) 
and the simulation code ZVAR that linearly
interpolates the evolutionary tracks both in age and metallicity,
including the effects of binary stars. The observational effects were
applied to the theoretical random populations using the photometric
errors and completeness obtained from the results of artificial star
experiments.

\realfigure{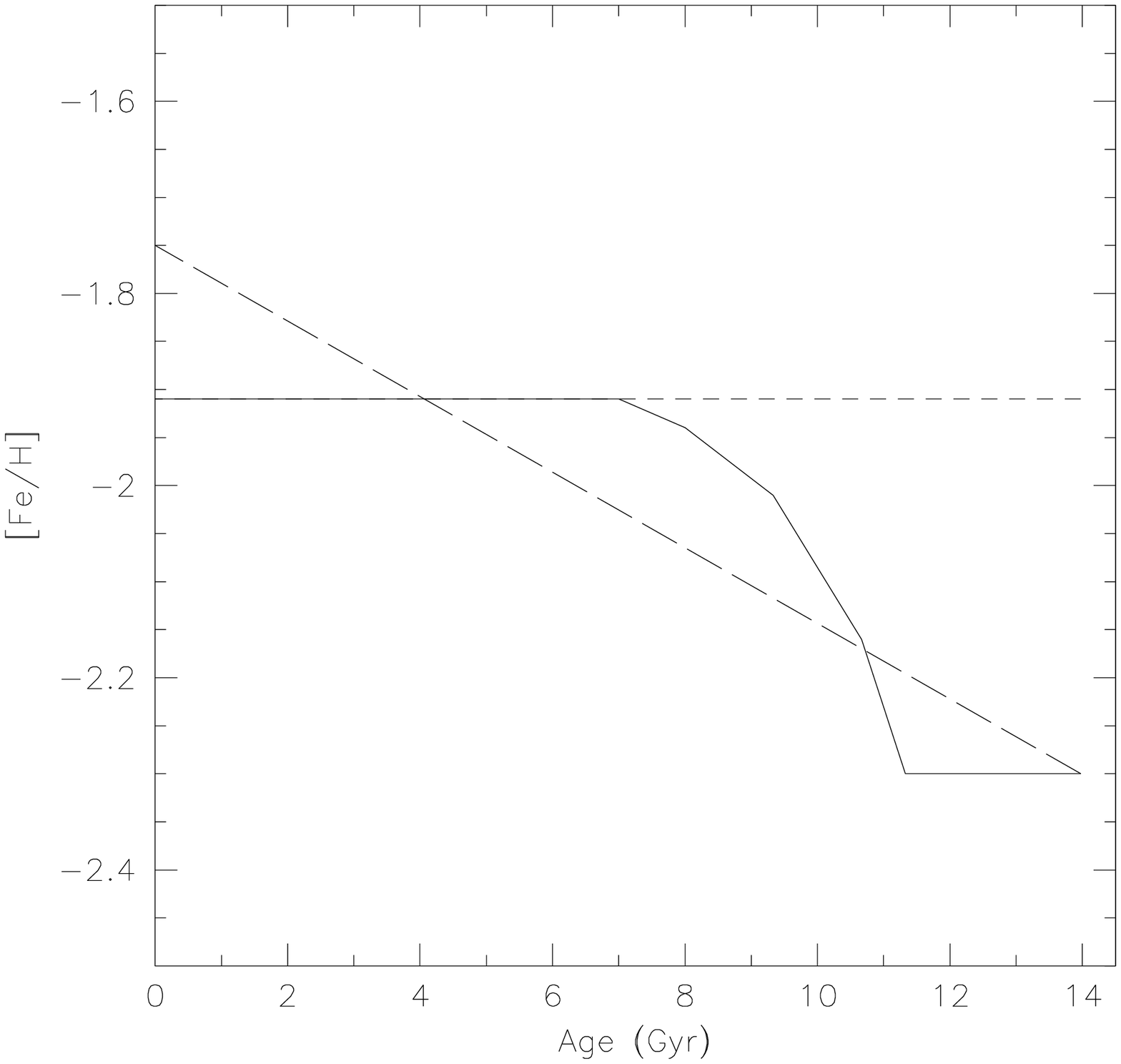}{
The different chemical-enrichment histories tested in this work. The
{\it continuous line} represents the adopted PT law, while the {\it
long-dashed} and {\it short-dashed} lines show the linear and constant
laws, respectively.  
}{f_leggi}

A basic ingredient of the simulation is the chemical-enrichment
history (\abbrev{CEH}), for which we chose to test a number of
reasonable assumptions (Fig.~\ref{f_leggi}).
{The first tested law is characterized by a prompt initial
enrichment at ages of $\sim 12$ Gyr, which is responsible for
producing much of the observed metals, followed by a slow metallicity
increase leading to a final value of $\feh \sim -1.9$ (hereafter ``PT
law'', after the age-metallicity relation modeled by Pagel \&
Tautvaisiene \cite{pag+tau98} for the Large Magellanic Cloud). 
} 
This trend is similar to the CEH recently derived from spectroscopic
observations of dwarf spheroidal galaxies (Gallart et
al. \cite{gal+02}; Tolstoy et al. \cite{tols+03}). The final
metallicity was constrained to yield {\it a mean metallicity over the
galaxy lifetime equal to the value derived from the photometry of
Carina.}
{This was done iteratively since the galaxy's star-formation
history is affected by the CEH normalization, and vice versa.}

Other tested laws, illustrated in Fig.~\ref{f_leggi}, include a 
constant metallicity law at $\feh=-1.91$ and a linear
law starting at $\feh=-2.3$ and reaching $\feh=-1.76$ at the present
epoch.  These laws sample a wide range of possibilities in the rate
of metal enrichment, still yielding the correct mean [Fe/H] over the
SFH.
%
{Other chemical evolution laws with a more metal-rich normalization 
would be inconsistent with the photometric data.}

A total of 7 populations were generated using these input laws, with
age bins centered at 1, 2.25, 4, 6, 8.5, 11 and 13 Gyr, each bin
containing 30.000 stars. Tests were made with both a 
30\% fraction of binary stars and no binaries at all. 
The SFH that best fits the observed color-magnitude diagram
was found by comparing the number of real and simulated stars inside a
number of regions in the CMD (``boxes'').  These were chosen to map
(i) the luminosity function along the main sequence, (ii) the relative
contribution of blue and red horizontal branch stars, (iii) the number of 
``red clump'' stars,
and (iv) the total number of stars along the RGB (see
Fig.~\ref{f_dati}, {left panel}).  
The width of the RGB itself is not an
input constraint for the best-fit CMD.
%
%

\realfigure{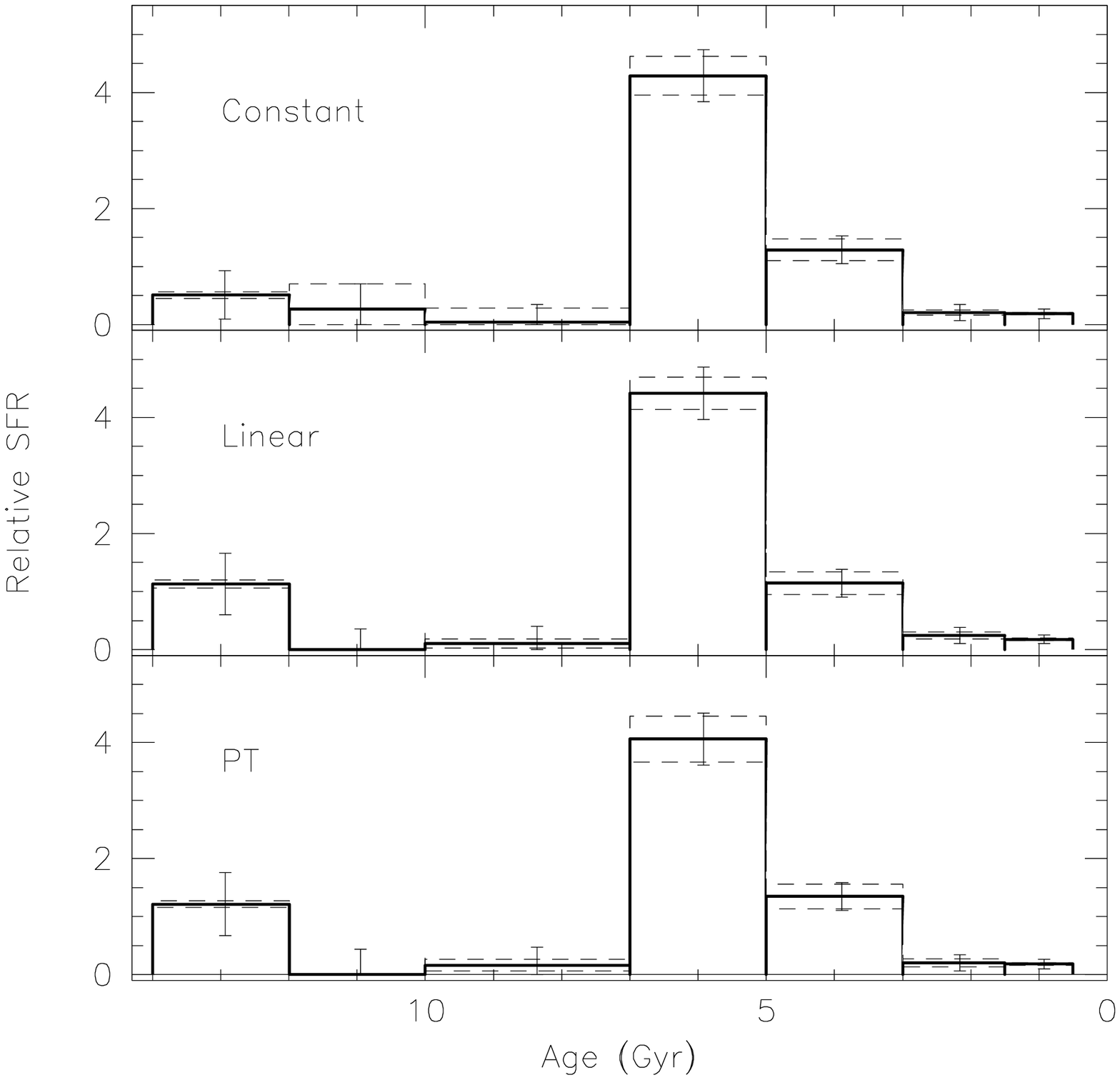}{
Relative star-formation history for the central region of Carina.  The
rates are normalized to a lifetime average rate of 25.6 $M_{\odot}$
Myr$^{-1}$. The error bars represent $3\sigma$ confidence intervals
from the chi-square statistic, while {\it dashed lines} show the upper
and lower limits to the SFH derived from 100 repeat CMD simulations.  The
bulk of the star formation appears to have happened in a long episode
between  $\sim 3$ and  $\sim 7$ Gyr ago.
}{f_sfhin}

The best fitting synthetic diagram, obtained by minimizing the
differences in star counts between the simulated and observed CMD with
a downhill $\chi^2$ minimization algorithm, is shown in the
right panel of Fig.~\ref{f_dati} for the PT law.

The derived star-formation histories are shown in Fig.~\ref{f_sfhin},
for the different assumptions on the metal-enrichment history.  Our
analysis of wide-field data agrees with previous results 
{in that $\sim 75$\% of the stars}  
formed between 3 and 7 Gyr ago, with a small yet
significant old stellar component. 
Note that, although the details of the SFH are different, the main
episodes of star formation are relatively independent of the adopted
CEH.  In fact, our suitable choice of the CMD regions over which stars
are counted makes the derived SFH more sensitive to the lifetimes of
evolutionary phases than to the metal enrichment law (and
uncertainties on the isochrone colors). 


\subsection{A thin RGB in Carina}
\label{thin}
We can now return to the problem of the origin of the extremely thin
RGB in Carina, having in hand the necessary tools to quantitatively
evaluate the effects of the chemical-enrichment and star-formation
history.
The color distribution of simulated RGB stars about a fiducial line
was measured for each choice of the CEH, using the same methods as for
the real data.  We did not make use of the absolute colors of the
synthetic RGBs, because of known uncertainties on the isochrone
colors.  The mean widths measured for the simulated diagrams are
$\sigma_{(V-I)} = 0.023 \pm 0.003$ for the linear and constant laws,
and $\sigma_{(V-I)} = 0.025 \pm 0.003$ for the PT law.  The mean
values and statistical errors result from 300 trials.
%
%
These results are essentially independent of the adopted fraction of
binary stars.
%
%
The simulations thus indicate that {\it the Carina's narrow RGB
naturally arises from the SFH derived from the data}.  The results for
all the tested enrichment laws agree with the observed color spread
within one standard deviation. Indeed, these values probably mark the
lower limit to the measurable RGB width set by stochastic effects and
photometric errors in the CMD simulations.
%
%
As expected, no conclusions can be reached on the details of chemical
enrichment from photometry alone. A large sample of spectroscopic
measurements will help, provided that it is of the highest precision
and accuracy. 


\section{Discussion and conclusions}

In this \letter, we have reported new measurements of the mean
metallicity ([Fe/H]=$-1.91 \pm 0.12$, after correction for the effects
of age) and the color spread ($\sigma_{V-I}=0.021 \pm 0.005$) of red
giant stars in Carina, based on new wide-field observations.
Using synthetic color-magnitude diagrams to model the color
distribution of RGB stars, we have quantitatively shown that the
thinness of the red giant branch is a direct consequence of the
derived star-formation history, for most chemical evolution laws that
are plausible and consistent with the observed mean metallicity of the
galaxy. 

The most fundamental reason for the RGB being so thin is the fact that
the predominant intermediate-age stellar population that makes up the
RGB, was formed in a limited time interval ($5 \pm 2$ Gyr ago), with
only a minor contribution by an old population.
In theory, both the age range and the metallicity evolution of the red
giant stars contribute in some way to the color range of the RGB, but
the effects are subtle and hardly measurable.
In fact, (i) the age interval is too small for the isochrone
separation to be measurable, and (ii) the sensitivity of the RGB color
to changes in [Fe/H] is low for metal-poor systems, so the effect of
{\it modest} metal enrichment during the main star-formation episode
is negligible too.

Our photometric analysis bears some interesting implications on the
chemical evolution history, by providing evidence that Carina could
not experience a significant metal enrichment (i.e. above [Fe/H]$\sim
-1.9$), certainly not before 3 Gyr ago. In fact, the lifetime-averaged 
metallicity inferred from photometric observations 
definitely rules out a metal-rich
final composition. Little information is available on the most recent
epoch, because stars younger than $\sim 3$ Gyr are only a trace
population, but for the same reason no significant metal enrichment is
expected in the last few Gyr. The most likely scenario remains that
indicated by chemical evolution models, in which substantial metal
enrichment is associated with the earliest star-formation epoch
($\gtrsim 12$ Gyr).

In summary, both the episodic star-formation activity (with
long-lasting episodes and quiescent phases) and the lack of a
significant metal enrichment suggest that metal-enriched gas outflows
play a role in the evolution of Carina and other low-mass
galaxies (Vader \cite{vade86}; Mac~Low \& Ferrara \cite{macl+ferr99}; 
see also Gallagher \& Wyse \cite{gal+wys94}; Kunth \& \"Ostlin
\cite{kun+ost00}; Carigi et al. \cite{cari+02}). 
Closed-box models
(e.g., Ikuta \& Arimoto \cite{ikut+arim02}) appear unable to explain
the very low metal abundances of dwarf galaxies like Carina.
In a similar way, the oxygen abundances measured in the interstellar
gas of dwarf irregular galaxies appear hardly consistent with a
closed-box model (see, e.g., Saviane et al. \cite{savi+02}, and
references therein).  
Observational evidence of galactic winds provides support to this
scenario {for active starbursts}, a recent example being
the Chandra observation of a metal-enriched galactic wind in the dwarf
galaxy NGC\,1569 (Martin, Kobulnicky, \& Heckman \cite{mart+02}).


\acknowledgments 

We thank the referee for suggesting some significant improvements to
the paper, Drs. L. Carigi and X. Hernadez for helpful remarks, and
M. Tosi for useful conversations on dwarf galaxy evolution. Support
for this work was provided by the Italian Ministry for University and
Research through grants COFIN2001028897 and COFIN2002028935.


\end{document}